\documentclass[authoryear,preprint,10pt,review]{elsarticle}
\usepackage{amssymb,amsmath,gensymb,multirow}
\usepackage{booktabs}
\usepackage{lineno}
\usepackage{float}
\usepackage{color}
\usepackage{algorithm,url}
\usepackage{algpseudocode}

\journal{arXiv} 

\begin{document}

\begin{frontmatter}

\title{Estimating Coverage in Streams via a Modified CVM Method}

\author[UC]{Carlos Hernandez-Suarez\corref{cor1}}
\ead{cmh1@cornell.edu}

\address[UC]{Coordinaci\'on General de Investigaci\'on Cient\'ifica, Universidad de Colima\\
Av. Gonzalo de Sandoval 444\\
Colima, Colima, 28045, MEXICO\\
cmh1@cornell.edu}
\date{\today}
\cortext[cor1]{Corresponding author}

\begin{abstract}
When individuals in a population can be classified in classes or categories, the coverage of a sample, $C$, is defined as the probability that a randomly selected individual from the population belongs to a class represented in the sample. Estimating coverage is challenging because $C$ is not a fixed population parameter, but a property of the sample, and the task becomes more complex when the number of classes is unknown. Furthermore, this problem has not been addressed in scenarios where data arrive as a stream, under the constraint that only $n$ elements can be stored at a time. In this paper, we propose a simple and efficient method to estimate $C$ in streaming settings, based on a straightforward modification of the CVM algorithm, which is commonly used to estimate the number of distinct elements in a data stream.
\end{abstract}


\end{frontmatter}

\section{Introduction}
Consider a data stream $\mathcal{A} = \{a_1, a_2, \ldots, a_m\}$ consisting of $m$ elements, each $a_i \in \mathcal{Y}$. Let $\mathcal{X} = \{x_1, x_2, \ldots, x_n\}$ be a subset drawn from $\mathcal{A}$ according to some sampling scheme. The \emph{coverage} $C$ is defined as the fraction of elements in $\mathcal{A}$ that appear at least once in $\mathcal{X}$. Equivalently, if an element is selected uniformly at random from $\mathcal{A}$, the probability that it also appears in $\mathcal{X}$ is $C$. For instance, suppose $\mathcal{A}$ represents the stream of requests to a digital service, and $\mathcal{X}$ is a sample of those requests, involving $k$ distinct users. A coverage value of $C$ then indicates that these $k$ unique users were responsible for $C \cdot 100\%$ of the total requests.

In streaming settings, we must process each element of $\mathcal{A}$ consecutively but can only store up to $n$ elements at a time (i.e., we can “remember” only $n$ elements). If the size $m$ of the stream is known in advance, selecting each element with probability $p = n/m$ yields a random sample of expected size $n$, allowing coverage estimation by existing methods. The problem becomes more difficult when $m$ is \emph{unknown}. Our goal is to develop a memory-efficient approach to estimate $C$ without prior knowledge of $m$. 

\section{The CVM Algorithm}
The CVM (Chakraborty, Vinodchandran, and Meel) algorithm is a classic method for approximating the number of distinct elements in a data stream \citep{chakraborty2023distinct}. It operates with limited memory by maintaining a fixed-size buffer $n$ and processing the stream in rounds:
\begin{itemize}
    \item In each round, new distinct elements from the stream are added to the buffer until it is filled, skipping duplicates that already exist in the buffer.
    \item Once the buffer is full, each element in the buffer is “flipped out” (removed) with probability $1/2$, leaving approximately half of the elements for the next round.
    \item The process continues in subsequent rounds with an increasing number of coin flips per element, making it progressively less likely that earlier elements remain.
    \item At the end of round $k$, each retained element corresponds to a probability of $\frac{1}{2^k}$ of surviving all coin flips. Thus, if there are $n$ elements in the buffer at that time, the estimated number of distinct elements is $n \cdot 2^k$.
\end{itemize}
Because it uses probabilistic halving to keep the buffer from overflowing, CVM is particularly suitable for settings in which memory is constrained and the stream size is large or unknown.

\section{Estimating Coverage with a Random Sample}
\citet{good1953population} first showed that for a sample of size $r$, where $s$ of those $r$ elements are \emph{singletons} (elements that appear exactly once), the quantity
\[
1 - \frac{s}{r}
\]
is a useful estimator of coverage. He credited this result to a personal communication with Turing. Later, \citet{Hernandez-Suarez2018} arrived at the same conclusion by modeling sampling from a multinomial distribution with an unknown number of categories: the key insight is that in a sample with singletons, the most likely scenario (when nothing is known about the total number of classes) is that each singleton belongs to its own unique class. Therefore, singletons do not contribute to coverage, implying that the coverage in the sample is primarily determined by elements observed more than once. Consequently, the estimated fraction of the population covered by the sample is $1 - \frac{s}{r}$, where $s$ is the count of singletons.

\section{Estimating Coverage in a Streaming Context}
To adapt Good’s estimator to streaming scenarios, we modify the CVM algorithm so that \emph{all} elements in $\mathcal{A}$ are tested for inclusion in $\mathcal{X}$ with equal probability in each round. Unlike the original CVM approach, we do not restrict ourselves to only adding elements that are not already in $\mathcal{X}$; instead, every incoming element has a chance to enter the sample. This ensures that each element in the stream has the same overall probability of being sampled, resulting in an unbiased random sample of size $n$ from the (potentially unknown) total stream size $m$. Once such a random sample is obtained, Good’s formula provides the coverage estimate:
\[
\widehat{C} \;=\; 1 - \frac{s}{n},
\]
where $s$ is the number of singletons in the final buffer $\mathcal{X}$. A sketch of the modified algorithm is shown below.

\begin{algorithm}
\caption{$C$-Estimator}
\begin{algorithmic}[1]
\State \textbf{Input:} Stream $A = (a_1, a_2, \ldots, a_m)$, integer $n$ (buffer size)
\State \textbf{Initialize:} $p \gets 1$, \quad $\mathcal{X} \gets \emptyset$
\For{$i = 1$ to $m$}
    \State \textbf{With probability } $p$, add $a_i$ to $\mathcal{X}$
    \If{$|\mathcal{X}| = n$}
        \State Remove each element of $\mathcal{X}$ with probability $1/2$
        \State $p \gets p/2$
        \If{$|\mathcal{X}| = n$}
            \State \textbf{Output} $\perp$ \quad (\emph{signal that the buffer is still full})
        \EndIf
    \EndIf
\EndFor
\State $s \gets \text{number of singletons in } \mathcal{X}$
\State \textbf{Output} $1-\frac{s}{n}$ \quad (\emph{coverage estimate})
\end{algorithmic}
\end{algorithm}

In this procedure, the probability $p$ of admitting new elements adjusts (halves) whenever the buffer reaches its capacity $n$. By flipping out half of the current elements in $\mathcal{X}$, we free space for continued sampling from the remainder of the stream. This ensures that, on average, each new element in the stream has the same chance of being included, creating a truly random sample of expected size $n$.

\section{Example: \emph{One Hundred Years of Solitude}}
We applied this method to the novel \emph{One Hundred Years of Solitude}, a landmark 1967 work by Colombian author Gabriel Garc\'ia M\'arquez. Widely hailed as a masterpiece of modern literature and a cornerstone of magical realism, the novel chronicles the multigenerational story of the Buend\'ia family in the fictional South American town of Macondo. We used the original version in Spanish, which contains 137{,}738 words, and estimated coverage at buffer sizes of $n = 100, 250, 500, 1000,$ and $2000$. For each buffer size, we performed $1000$ simulations and calculated the difference between the estimate and the true coverage. The results are shown in Figure~\ref{fig:Rplot}.

\begin{figure}[htbp]
    \centering
    \includegraphics[width=1\textwidth]{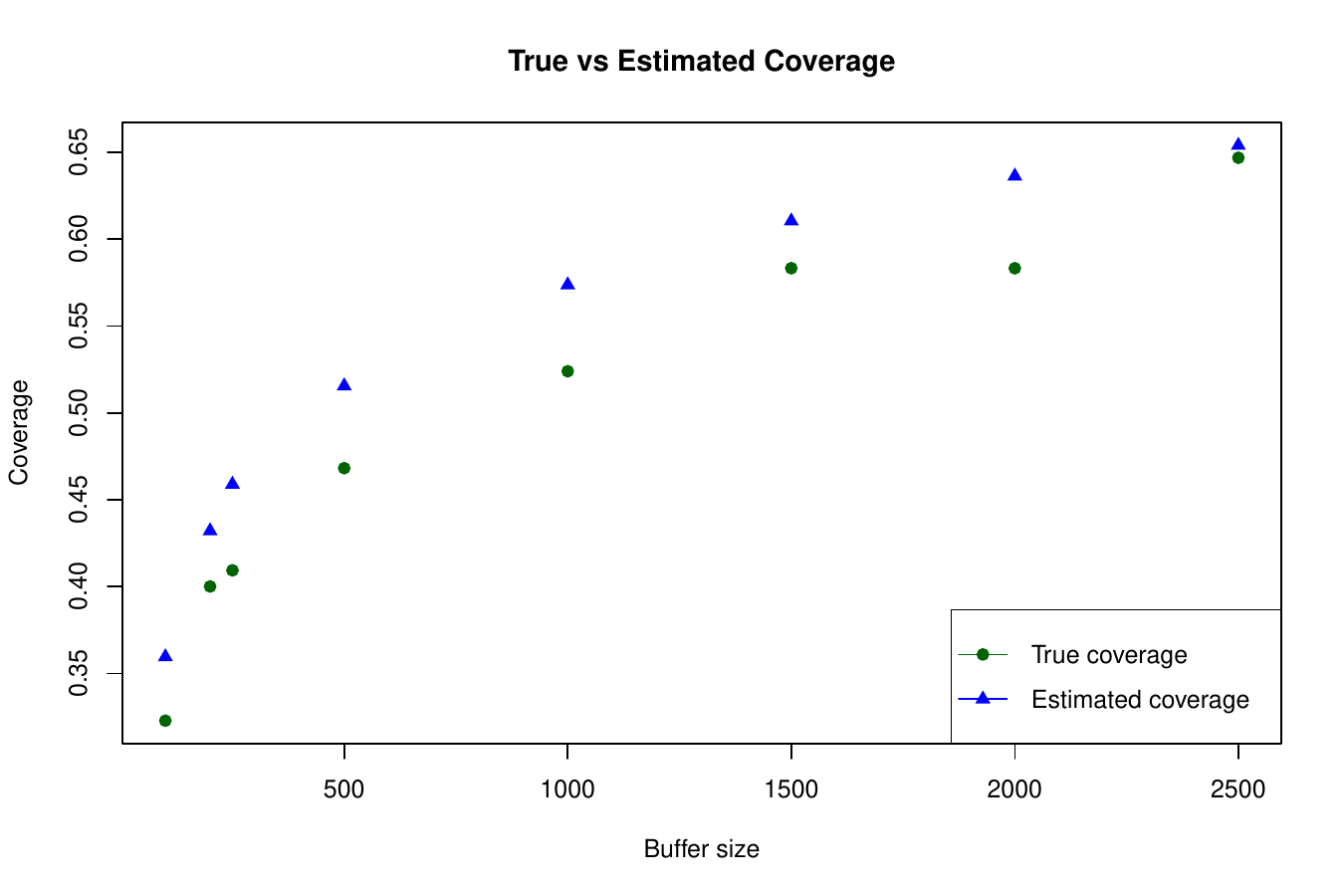}
    \caption{True vs. estimated coverage at different buffer sizes. Results of 1,000 simulations.}
    \label{fig:Rplot}
\end{figure}

\begin{figure}[htbp]
    \centering
    \includegraphics[width=1\textwidth]{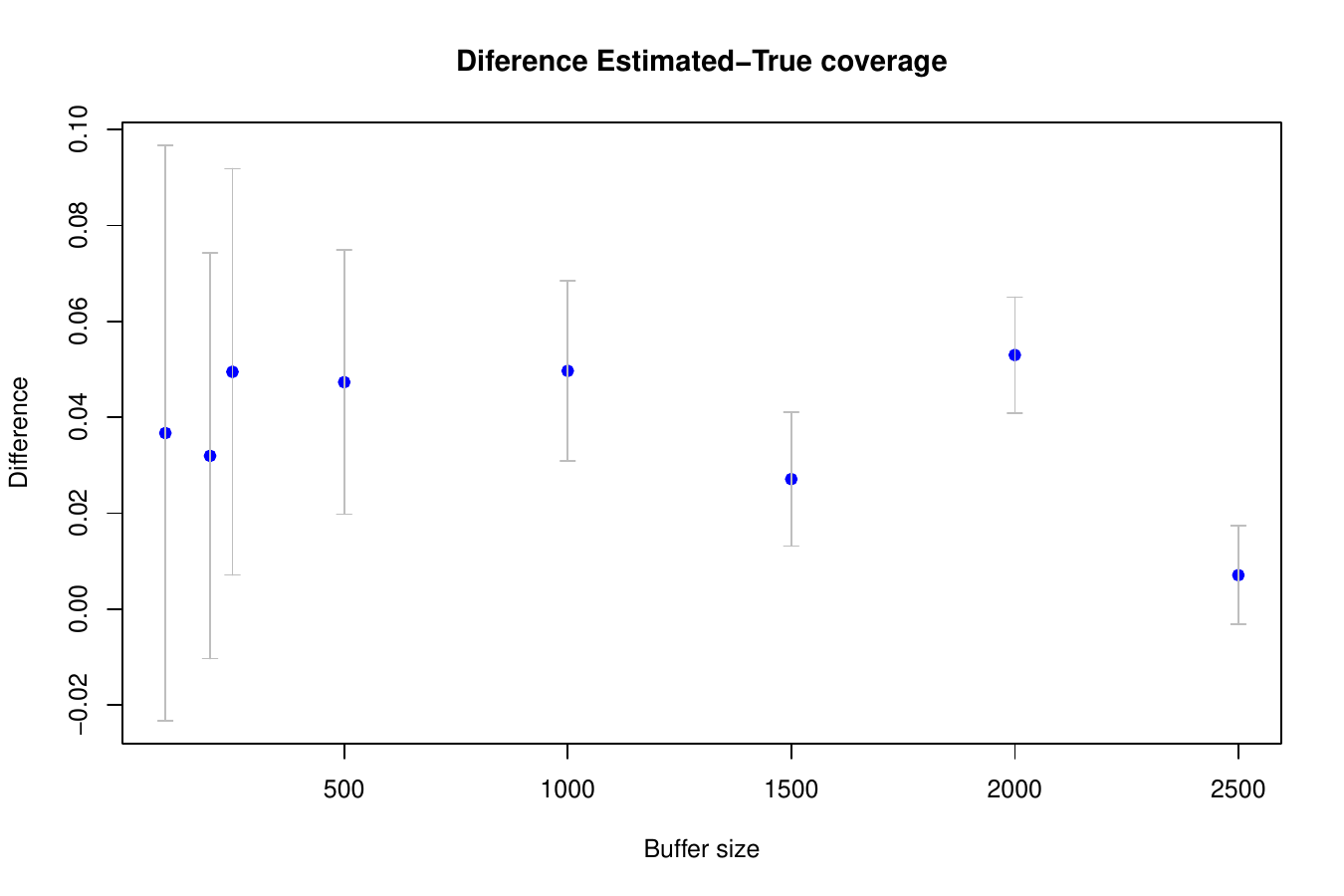}
    \caption{Observed difference between estimated and true coverage at different buffer sizes. Results of 1,000 simulations. Vertical lines represent error bars.}
    \label{fig:Rplot}
\end{figure}

\section{Discussion}
As shown in Figure~\ref{fig:Rplot}, the difference between the estimated coverage and the true coverage remains below 0.06 across all tested buffer sizes. We also observe that the standard deviation of the estimates decreases as the buffer size increases, indicating that larger buffers yield more precise estimates. This behavior is expected, since a larger buffer is more likely to produce a representative sample of the underlying data stream, thereby reducing variability in the estimation process.

We have demonstrated how a simple, memory-efficient adaptation of the CVM algorithm can generate a random sample from a potentially unbounded or unknown-size data stream. Once the sample is obtained, coverage can be estimated via Good’s singletons-based formula, \(1 - \frac{s}{n}\). This modification provides a practical solution for streaming contexts in which storing or processing all data is infeasible, enabling real-time coverage estimation with limited memory.


\section{Bibliography}

\begin{thebibliography}{3}
\providecommand{\natexlab}[1]{#1}

\bibitem[{Chakraborty \emph{et~al.}(2023)Chakraborty, Vinodchandran \&
  Meel}]{chakraborty2023distinct}
Chakraborty, S., Vinodchandran, N. \& Meel, K.S. (2023) Distinct elements in
  streams: An algorithm for the (text) book.
\newblock \emph{arXiv preprint arXiv:230110191}.

\bibitem[{Good(1953)}]{good1953population}
Good, I. (1953) The population frequencies of species and the estimation of
  population parameters.
\newblock \emph{Biometrika}, \textbf{40}, 237--264.

\bibitem[{Hernandez-Suarez(2018)}]{Hernandez-Suarez2018}
Hernandez-Suarez, C. (2018) Measuring the representativeness of a germplasm
  collection.
\newblock \emph{Biodiversity and Conservation}, \textbf{27}, 1471--1486.
\newblock \url{https://dx.doi.org/10.1007/s10531-018-1504-3}.
\end{thebibliography}
\end{document}